\documentclass[aip,pof,amsmath,amssymb,reprint]{revtex4-2}
\usepackage{booktabs}
\usepackage{graphicx}
\usepackage{bm}
\usepackage{url}

\begin{document}

\title{Lagrangian transport driven by travelling waves in a radially deforming circular cavity}

\author{Zijian~Liu}
\thanks{Z. Liu and J. Olszewski contributed equally to this work.}
\affiliation{Department of Mechanical and Industrial Engineering,
University of Illinois Chicago, Chicago, Illinois 60607, USA}

\author{Julian~Olszewski}
\affiliation{Department of Mechanical and Industrial Engineering,
University of Illinois Chicago, Chicago, Illinois 60607, USA}

\author{Jie~Xu}
\email[Author to whom correspondence should be addressed: ]{jiexu@uic.edu}
\affiliation{Department of Mechanical and Industrial Engineering,
University of Illinois Chicago, Chicago, Illinois 60607, USA}

\date[]{}

\begin{abstract}
A travelling deformation of a circular cavity wall can carry particles with the wave or
against it. For two counter-propagating waves of azimuthal order $m$ and amplitudes $a$
and $b$, the leading axisymmetric component of the total (Lagrangian) drift scales with
the intensity difference $S_3=|a|^2-|b|^2$. We obtain the radial profile of the azimuthal
drift in closed form in the viscous limit, where the Womersley number $\mathrm{Wo}$ tends
to zero, and by quadrature for any finite $\mathrm{Wo}>0$.
The total drift is the sum of the Eulerian mean velocity and the Stokes drift.
For radial wall motion with $m\ge2$ and unequal wave intensities, the total drift follows
the stronger of the two waves throughout the interior in the viscous limit, although the
Stokes drift changes sign with radius for all $\mathrm{Wo}$.
For $m=2$ and large $\mathrm{Wo}$, the predicted total drift just inside a radially moving
wall opposes the wave to leading order in wall amplitude. For the same travelling shape,
the total drift there follows the wave when the wall material also moves tangentially to
preserve local arc length to first order.
For the radial wall at large $\mathrm{Wo}$, the near-wall profile consists of the planar
Longuet-Higgins result and a closed-form curvature correction. At large $m$ the core
rotation approaches Tuck's finite-Reynolds-number waving-sheet result.
At $m=2$ and $\mathrm{Wo}^2=10$, moving-domain simulations recover the leading-order
drift coefficients for both wall motions.

\end{abstract}

\maketitle

\section{Introduction}
\label{sec:intro}

Travelling waves on the wall of a compliant chamber or peristaltic mixer can rotate its
contents. Attenuation in surface acoustic wave devices provides one example of how
travelling components of unequal strength can arise even in a nominally standing
wave~\cite{devendran2016}. For a circular cavity driven by prescribed wall
motion, we ask where particles drift with the stronger wave and where they drift against it.
Oscillatory forcing drives steady streaming through Reynolds stresses in the bulk and in
viscous boundary layers~\cite{riley2001,boluriaan2003}. For the total drift we use
the second-order decomposition of generalized Lagrangian-mean theory~\cite{andrews1978},
\begin{equation}
U_L = U_E + U_S ,
\label{eq:decomp}
\end{equation}
where $U_L$ is the leading Lagrangian drift, $U_E$ the Eulerian mean velocity, and $U_S$ the Stokes drift.
We average each azimuthal component over a period and a reference ring in the undeformed disk.
In the interior that remains fluid throughout the cycle, the Eulerian mean is evaluated at fixed physical points.
The Stokes drift accounts for the correlation
between the oscillatory particle displacement and the spatial gradient of the oscillatory velocity. We distinguish local fields from their azimuthal averages throughout.

\emph{Prior chirality-dependent streaming.} Riley's cylinder in orthogonal sound beams,
equivalently an elliptically orbiting cylinder, has an
axisymmetric slip carried by its minor-to-major axis ratio $\lambda$~\cite{riley1992,riley1998}.
With unit major-axis amplitude, Riley's axis ratio equals the difference in intensity between
the two counter-rotating components. His exterior thin-layer slip and the drift inside the
cavity use different flow and orientation conventions. Lee and Wang's streaming and torque for two orthogonal standing waves about a sphere contain
terms proportional to the product of the two amplitudes and the sine of their phase
difference~\cite{leewang1989}. Rednikov and Sadhal's high-frequency effective slip likewise
contains a rotational contribution set by the cross product of the incident velocity
quadratures~\cite{rednikov2011}. These measures are proportional to a circular-polarization
intensity imbalance after fixing the normalization.
The present calculation determines the radial coefficient for a deforming circular interior,
including finite-mode curvature and finite Womersley number.

\emph{Prior Lagrangian transport in closed containers.} Selverov and Stone compute the
Eulerian mean flow in a rectangular channel driven by a transverse travelling wall wave at
arbitrary frequency, and compare Eulerian and Lagrangian means through high- and
low-frequency asymptotics~\cite{selverov2001}. Their channel is closed at both ends, so the
mean flow carries no net flux through any cross section and transport near the walls is
balanced by a return flow; at low frequency their near-wall Lagrangian transport opposes the
wave at most positions along the channel. Azimuthal circulation in the disk can have one sign
throughout the interior, since its geometry permits a net flux through a radial cut.
Yi et al. treat a closed channel with two vibrating walls~\cite{yibauhu2002}.
Ng and Ma obtain one-dimensional and cellular mean transport from waves on the two channel
walls~\cite{ngma2009}. Ilin and Morgulis obtained high-frequency Lagrangian transport between two
vibrating planes~\cite{ilin2012}. For a circular pipe driven by a helical wave of radial wall
motion, they found that particles at different radii can rotate in opposite directions~\cite{ilin2019}.
In the disk, the wall waves travelling towards increasing and decreasing azimuth
have complex amplitudes $a$ and $b$, respectively. The axisymmetric drift is proportional to their
intensity difference $S_3=|a|^2-|b|^2$.
In a partially filled, orbitally shaken cylinder, Bouvard et al. measure Lagrangian mean
rotation whose core is consistent with the inviscid Stokes drift of the rotating sloshing
wave~\cite{bouvard2017}. Here the fluid fills the disk, and the wall material can move
tangentially along its prescribed shape.

\emph{Planar and rotating-container counterparts.} In the Stokes limit, Taylor's transversely
waving sheet drives a flow $\tfrac12\omega kb^2$ at infinity relative to the sheet, for wave
number $k$ and amplitude $b$~\cite{taylor1951}; Tuck extended the calculation to finite
Reynolds number $\omega/(\nu k^2)$, at which the speed falls towards half that
value~\cite{tuck1968}. Under a progressive wave, the oscillatory boundary layer on a plane bed
carries a mass transport $\tfrac54U^2/c$ at its edge, for slip-velocity amplitude $U$ and phase
speed $c$~\cite{longuethiggins1953}. Fung and Yih computed peristaltic transport by a travelling
channel-wall wave at finite Reynolds number in a small-amplitude expansion~\cite{fungyih1968}.
Shapiro and Jaffrin examined the distinction between Eulerian and Lagrangian means in
identifying reflux~\cite{shapirojaffrin1971}. In the
Lagrangian-mean treatment of streaming by leaky surface acoustic waves by Vanneste and B\"uhler, the
slip velocity includes the oscillation of the boundary, and the mean flow of the particles separates
into Eulerian mean flows driven in the interior and at the boundary, and the Stokes
drift~\cite{vannestebuhler2011}. Outside a circular
cylinder, small surface oscillations propel the cylinder in the Stokes limit~\cite{blake1971}.
In a rapidly rotating cylinder, a rotating multipolar wall deformation drives an axisymmetric
zonal flow through the Ekman layers on its end plates~\cite{sauret2015}; that mechanism
requires the rotation of the container and is absent here. The circular interior studied below
recovers the planar results at large $m$, and its finite-$m$ differences from them are
curvature terms.

\emph{Prior streaming in deformable containers and interfaces.} Experiments on an oscillating
deformable cylinder relate interior streaming to dimensionless frequency~\cite{kozlov2017}.
Gritsenko et al. identified azimuthal modes of a pinned air--water interface
oscillating out of the plane of a circular microcavity opening~\cite{gritsenko2018}.
Drop calculations treat oscillatory interfaces with an adsorption layer~\cite{murtsovkin1992},
and arbitrary axisymmetric shape oscillations of a fluid particle~\cite{doinikov2026}.
These provide context for the role of wall motion and penetration depth.
In a villi-patterned channel, phase-lagged gaps produce transport opposite to a pendular
wave~\cite{vernekar2025}.
Elastic-tube rectification couples prescribed forcing to wall compliance~\cite{zhang2024};
the present wall motion is specified kinematically.

The present result determines the leading axisymmetric component of the total drift for
counter-propagating waves of one azimuthal order in a radially deforming circular cavity.
The standing-wave solution for this geometry already supplies the first-order Bessel profile,
transferred wall condition, the axisymmetric Reynolds shear of co-phased first-order factors
and the viscous-limit cellular Lagrangian mean~\cite{liu2026}.
Here the wave-intensity difference sets the amplitude of the axisymmetric total drift. A radial
quadrature gives its profile at finite Womersley number and reduces to a closed form in
the viscous limit. For radially moving wall material, the Stokes component can oppose the
wave on rings where the particles still follow it. We determine where finite inertia also
reverses the total drift. Holding the travelling shape fixed then lets us test which of
these reversals persist when the wall material moves along it.

\section{Problem}
\label{sec:problem}

An incompressible Newtonian fluid of kinematic viscosity $\nu$ fills a circular cavity of undeformed radius $R_0$.
The analysis and simulations assume axial invariance, as in the interior of a deep cavity far from its end walls.
We prescribe a wall disturbance
of azimuthal order $m\ge1$, angular frequency $\omega$ and small amplitude $\varepsilon$.
Incompressibility fixes the enclosed area and excludes the axisymmetric breathing mode ($m=0$). Lengths are scaled on $R_0$ and time on $1/\omega$, so
the Womersley number satisfies $\mathrm{Wo}^2 = \omega R_0^2/\nu$.
At fixed $R_0$ and $\nu$, increasing the driving frequency increases $\mathrm{Wo}$.
At small $\mathrm{Wo}$ the first-order vorticity fills the cavity, and at
large $\mathrm{Wo}$ it is confined near the wall, with dimensional penetration depth $\sqrt{2\nu/\omega}$ and thickness $\sqrt{2}/\mathrm{Wo}$ on the cavity-radius scale.
Second-order mean velocities are reported on the scale $\varepsilon^2\omega R_0$.
The expansion takes $\varepsilon\to0$ at fixed $m$ and
$\mathrm{Wo}$. Wall slope requires $\varepsilon m\ll1$, and displacement relative to the
Stokes-layer thickness requires $\varepsilon\mathrm{Wo}\ll1$ in the thin-layer regime.

\emph{Relation to acoustic actuation.} The incompressible model describes the response to a
prescribed wall motion. For acoustic actuation, the cavity must be compact
($\omega R_0/c_0\ll1$, with $c_0$ the sound speed) and the forcing must produce the prescribed wall motion.
A gas can be treated in this limit when density variations are negligible and the continuum
no-slip condition applies. Consider a water-filled cavity of radius $100\,\mu\mathrm{m}$ with
$\nu\approx10^{-6}\,\mathrm{m^2\,s^{-1}}$ and $c_0\approx1.5\times10^3\,\mathrm{m\,s^{-1}}$.
A drive at about $160\,\mathrm{Hz}$ gives $\mathrm{Wo}^2=10$ and a compactness parameter of about $7\times10^{-5}$.
For a radial travelling wave with $m=2$ and $S_3=1$, a wall displacement of $2\,\mu\mathrm{m}$
($\varepsilon=0.02$) gives a predicted total drift of about $66\,\mu\mathrm{m\,s^{-1}}$ at $r=0.45$.

At the $99.85\,\mathrm{MHz}$ operating frequency of the surface acoustic wave device
of Devendran et al.~\cite{devendran2016}, the same water-filled cavity would have
$\omega R_0/c_0\simeq42$, well outside the acoustically compact regime.

\begin{figure*}[t]
\centering
\includegraphics[width=\textwidth]{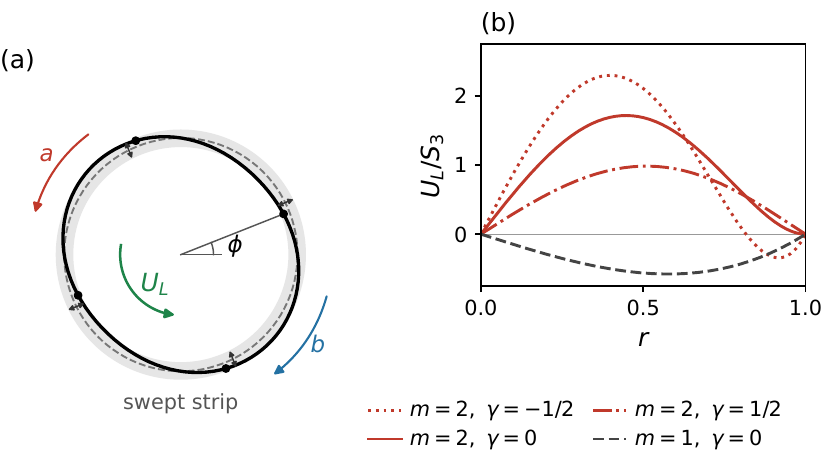}
\caption{Geometry and the effect of tangential wall motion. (a) The cavity for $m=2$, drawn
with an exaggerated amplitude $\varepsilon=0.075$: the dashed circle is the reference wall $r=1$,
the solid curve the wall at one instant and the shaded annulus the strip swept by the wall.
Wall material points (dots) keep their angular labels $\phi$ and move radially (double arrows beside them).
The $a$-wave travels towards increasing $\theta$ and the $b$-wave towards decreasing $\theta$.
In the viscous limit the interior particles follow the $a$-wave for this radial wall when the intensity difference $S_3=|a|^2-|b|^2$ is positive. (b) The
viscous-limit total drift of Eq.~\eqref{eq:gammaG} for $m=2$ with tangential-to-radial displacement
amplitude ratio $\gamma=-1/2$, $0$ (radial wall) and $1/2$ (first-order inextensible wall), and for $m=1$
with $\gamma=0$. All curves are leading-order coefficients on the reference disk.}
\label{fig:geometry}
\end{figure*}

\emph{The wall motion.} For the radial wall, material points keep fixed angular coordinates
$\phi$ [Fig.~\ref{fig:geometry}(a)]. For a single travelling wave the dimensionless wall radius $R(\phi,\tau)$ satisfies
$R^2=1-2\varepsilon\sin(m\phi-\tau)$. The enclosed area is $\pi$ at every instant.
The wall points have zero azimuthal velocity, including for $m=1$, and their radial motion generally
has a component along the deformed curve. We also consider material points that move
tangentially along the same prescribed shape (Secs.~\ref{sec:total}--\ref{sec:verification}).

\emph{Harmonic time dependence and the first-order field.} Every first-order quantity is written as
$\mathrm{Re}\{\,\cdot\,e^{-i\tau}\}$ with $\tau=\omega t$. Resolving the wall motion into two
counter-propagating components of complex amplitudes $a$ and $b$, the first-order streamfunction is
\begin{equation}
\psi_1 = \mathrm{Re}\!\left\{f(r)\left[a\,e^{im\theta} + \bar b\,e^{-im\theta}\right]e^{-i\tau}
\right\} ,
\label{eq:psi1}
\end{equation}
which is $\mathrm{Re}[f a\,e^{i(m\theta-\tau)}] + \mathrm{Re}[\bar f b\,e^{i(m\theta+\tau)}]$: the
$a$-wave travels towards increasing $\theta$ and the $b$-wave towards decreasing $\theta$. The
second term carries the conjugate profile $\bar f$, as required by the vorticity equation. Writing
$L_m = \mathrm{d}^2/\mathrm{d}r^2 + r^{-1}\mathrm{d}/\mathrm{d}r - m^2/r^2$ for the radial Laplacian
at order $m$, the linearized vorticity equation
$\mathrm{Wo}^2\,\partial_\tau\nabla^2\psi_1 = \nabla^4\psi_1$ requires
\begin{equation}
\left(L_m + i\,\mathrm{Wo}^2\right)L_m f = 0
\label{eq:ode}
\end{equation}
for the $e^{-i\tau}$ profile, and the same equation with the opposite sign of
$i\,\mathrm{Wo}^2$ for the $e^{+i\tau}$ profile. The conjugate profile therefore carries the $b$-wave. Real equal amplitudes $a=b$ give the standing mode
$\mathrm{Re}[2a f\cos(m\theta)e^{-i\tau}]$, and $b=0$ a single travelling wave. Equal magnitudes
with complex phases give a standing mode with shifted spatial and temporal origins.

\emph{Boundary conditions.} $f$ is regular at the origin, where $f\sim r^m$, and at the wall
\begin{equation}
m f(1) = 1 , \qquad f'(1) = 0 .
\label{eq:bc}
\end{equation}
With $u_{1r}=r^{-1}\partial_\theta\psi_1$, the first fixes the radial velocity coefficient at
the reference wall to $\mathrm{Re}\{i[a e^{im\theta}-\bar b e^{-im\theta}]e^{-i\tau}\}$.
The physical first-order velocity is $\varepsilon\omega R_0$ times this coefficient: $a$ and $b$
are order-one amplitudes. The stated travelling material wall corresponds to $a=-i$, $b=0$,
so its radial velocity is $\varepsilon\cos(m\phi-\tau)+O(\varepsilon^2)$ in dimensionless units.
The phase changes neither $S_3$ nor the ring-averaged total drift.

The second condition, $f'(1)=0$, expresses the radial material-wall kinematics as a condition on
$f$: the azimuthal velocity at the wall is $-f'(1)\,\mathrm{Re}\{\cdots\}$, so the prescribed zero
azimuthal material velocity fixes $f'(1)=0$. The boundary condition determines the interior
flow: the uniform $m=1$ solution $f=r$, with $f'(1)=1$, has zero mean flow, while the radially
constrained wall gives the profile below. Equation~\eqref{eq:ode} with
Eq.~\eqref{eq:bc} gives the standing-wave radial solution~\cite{liu2026}, $f = Cr^m + B_m J_m(\kappa r)$ with $\kappa^2 = i\,\mathrm{Wo}^2$, and in
the viscous limit the elementary polynomial
\begin{equation}
f_0(r) = \frac{(m+2)r^m - m\,r^{m+2}}{2m} .
\label{eq:f0}
\end{equation}

\emph{Chirality.} Following the polarization convention of optics, define
\begin{equation}
S_0 = |a|^2+|b|^2, \qquad S_3 = |a|^2-|b|^2, \qquad \chi = S_3/S_0 ,
\label{eq:stokes}
\end{equation}
so $\chi=0$ is a standing mode and $\chi=\pm1$ a pure travelling wave. With unit major-axis
amplitude, Riley's ellipse has counter-rotating amplitudes $(1+\lambda)/2$ and $(1-\lambda)/2$.
Their intensity difference is $S_3=\lambda$, giving
\begin{equation}
\chi = \frac{2\lambda}{1+\lambda^2},
\label{eq:lambda}
\end{equation}
which is $0$ at $\lambda=0$, his transverse vibration, and $1$ at $\lambda=1$, his purely orbital
motion. Equation~\eqref{eq:lambda} relates the two descriptions of the drive.
The radial material-wall conditions in Eq.~\eqref{eq:bc} determine the flow inside the cavity.

\section{Stokes drift from the radial intensity and phase}
\label{sec:functional}

Define $\bm{\xi}_1(r,\theta,\tau)$ on the undeformed disk by
$\partial_\tau\bm{\xi}_1=\bm{u}_1$ and zero temporal mean. It is the first-order displacement
field evaluated at a fixed reference position. The second-order Stokes coefficient is
$\bm{u}_S=\langle(\bm{\xi}_1\cdot\nabla)\bm{u}_1\rangle_\tau$, with the polar basis derivatives
included and the average taken over a period. For Eq.~\eqref{eq:psi1} this gives
\begin{align}
u_{Sr} &= -2m^2\,\mathrm{Re}\!\left(ab\,e^{2im\theta}\right)\frac{W(r)}{r^2} ,
\label{eq:full_r}\\
u_{S\theta} &= S_3\,G(r)
 + m\,\mathrm{Im}\!\left(ab\,e^{2im\theta}\right)\frac{\mathrm{d}}{\mathrm{d}r}
 \!\left[\frac{W(r)}{r}\right] ,
\label{eq:full_t}
\end{align}
where
\begin{equation}
G(r) = \frac{m}{4}\,\frac{\mathrm{d}}{\mathrm{d}r}\!
\left[\frac{1}{r}\frac{\mathrm{d}|f|^2}{\mathrm{d}r}\right] ,
\qquad
W(r) = \mathrm{Im}\!\left(\bar f f'\right) = |f|^2\frac{\mathrm{d}\arg f}{\mathrm{d}r} .
\label{eq:GW}
\end{equation}
$G$ is set by how the first-order intensity is distributed in radius and $W$ by its radial phase
gradient. The phase-gradient function also appears in the displacement correction for the standing wave~\cite{liu2026}. The two terms carried by $ab\,e^{2im\theta}$ are pure $\cos2m\theta$ and $\sin2m\theta$,
so they drop out of the azimuthal average and
\begin{equation}
\langle u_{S\theta}\rangle_\theta = S_3\,G(r) ,
\qquad
\langle u_{Sr}\rangle_\theta = 0 .
\label{eq:main}
\end{equation}

In Eq.~\eqref{eq:main}, the intensity difference $S_3$ sets the amplitude of the ring-averaged Stokes drift.
The radial factor $G(r)$ depends on $|f|^2$ and is independent of the wave amplitudes and their phases.
At fixed total intensity $S_0$, the drift is therefore proportional to $\chi$ and vanishes for waves of equal intensity.
The result holds as $\varepsilon\to0$ at each fixed $\mathrm{Wo}$ for any first-order profile $f$ satisfying
Eq.~\eqref{eq:ode}, including the viscous-limit polynomial in Eq.~\eqref{eq:f0} and the Bessel profile
at finite $\mathrm{Wo}$. The finite-amplitude conditions are given in Sec.~\ref{sec:problem}.

At finite $\mathrm{Wo}$, the radial phase variation in the Stokes layer enters the local $2m$ harmonic through
$W$ in Eqs.~\eqref{eq:full_r} and~\eqref{eq:full_t}. For a standing wave, this harmonic has the angular
dependence of the $4m$-cell pattern~\cite{liu2026}, even though the drift averages to zero on each ring.
At fixed wave amplitudes the harmonic is $O(\mathrm{Wo}^2)$ in the viscous limit, since $W=O(\mathrm{Wo}^2)$.
For a single travelling wave ($ab=0$), or for a profile real up to a constant phase ($W\equiv0$),
the harmonic vanishes identically. Equation~\eqref{eq:main} then gives the drift at every point,
with $u_{Sr}=0$ and $u_{S\theta}=S_3G(r)$.

In the viscous limit Eq.~\eqref{eq:main} becomes elementary,
\begin{equation}
\langle u_{S\theta}\rangle_\theta = S_3\,\frac{(m+2)\,r^{2m-3}}{4}
\left[m(m+1)(r^2-1)^2-2\right] ,
\label{eq:elementary}
\end{equation}
which is regular at the origin for every $m\ge1$; at $m=1$ it is $\tfrac{3}{2}S_3\,r(r^2-2)$. At the
reference wall Eq.~\eqref{eq:elementary} is $-S_3(m+2)/2$. For $S_3>0$ the dominant wave
travels towards increasing $\theta$: the inner Stokes component follows that direction for $m\ge2$
and the outer component opposes it. This sign follows from the radial no-slip kinematics:
$f_0'(1)=0$ and $f_0''(1)=-(m+2)$ give $G(1)=f_0''(1)/2<0$. A radial displacement samples the
near-wall tangential velocity gradient, although each wall particle has zero tangential motion.
The Stokes drift at the reference wall is cancelled by the Eulerian mean, as derived below.
The coefficients are expansions on the reference disk. Boundary transfer supplies their
interpretation in the strip swept by the physical wall; the fixed-point fluid interpretation
applies to the interior region that remains fluid throughout the cycle.

\section{The complete leading axisymmetric transport}
\label{sec:total}

Write $U_E=\langle u_{E\theta}\rangle_\theta$ and similarly for $U_L$ and $U_S$, on the common
second-order scale. Since the material wall returns to its initial position every period,
its net particle displacement over each period is exactly zero. First-order no slip makes the wall
displacement equal to $\bm{\xi}_1(1)$, with zero tangential component. The averaged second-order
tangential no-slip condition transferred to the reference circle $r=1$, as in the standing-wave
calculation~\cite{liu2026}, is
\begin{equation}
\left\langle u_{2\theta}+\big[(\bm{\xi}_1\cdot\nabla)\bm{u}_1\big]_\theta\right\rangle_\tau(1)=0,
\label{eq:walltransfer}
\end{equation}
as the material wall has zero tangential velocity at every order. The mean radial
second-order displacement generated by the squared-radius prescription multiplies the zero
base-flow gradient and adds no tangential term here. Taking the azimuthal average gives
\begin{equation}
U_L(1)=0,\qquad U_E(1)=-U_S(1)=-S_3G(1).
\label{eq:meanwall}
\end{equation}
At $r=1$, $f'(1)=0$ gives $W(1)=0$, so the radial Stokes term vanishes. The tangential
non-axisymmetric term can remain nonzero at finite Womersley number, through $W'(1)$, and is
cancelled by the corresponding Eulerian boundary term. Equation~\eqref{eq:meanwall} states the
axisymmetric projection of that cancellation.

The axisymmetric selection also determines the leading total drift at finite $\mathrm{Wo}$.
The first-order response is linear in the drive, so its second-order mean forcing, boundary data
and Stokes drift are quadratic in $a$ and $b$. Period averaging and rotation invariance leave only
$|a|^2$ and $|b|^2$ in an axisymmetric quantity: the mixed term $ab$ carries angular order $2m$.
Reflection $\theta\mapsto-\theta$ interchanges the two intensities and reverses azimuthal velocity.
Their coefficients must therefore be equal and opposite. For the steady mean selected after
transients, with no independently imposed mean wall rotation,
\begin{equation}
U_E=S_3 E_m(r;\mathrm{Wo}),\qquad
U_L=S_3\left[E_m(r;\mathrm{Wo})+G(r)\right].
\label{eq:totalsymmetry}
\end{equation}
Any linear radial integral of this azimuthal transport inherits the same proportionality.
This symmetry result fixes the amplitude and phase dependence at each fixed $\mathrm{Wo}$
within the leading expansion. The second-order momentum equation determines $E_m$ and the
sign of the total drift.

The mean Reynolds stress per unit chirality is
$\mathcal R_m=\langle u_{1r}u_{1\theta}\rangle_{\tau,\theta}/S_3=-mW/(2r)$,
the axisymmetric Reynolds shear of co-phased first-order factors obtained in the standing-wave
calculation~\cite{liu2026}; the coefficient is defined by continuity when $S_3=0$.
Conservative azimuthal momentum gives
$E_m''+E_m'/r-E_m/r^2=\mathrm{Wo}^2(r^2\mathcal R_m)'/r^2$.
Writing $E_m=r\Omega_m$ and imposing regularity gives
$\Omega_m'=\mathrm{Wo}^2\mathcal R_m/r$.
The transferred wall condition in Eq.~\eqref{eq:meanwall} then fixes the unique regular solution:
\begin{align}
E_m(r;\mathrm{Wo})&=-rG(1)
 +\frac{m\mathrm{Wo}^2r}{2}\int_r^1\frac{W(s)}{s^2}\,\mathrm{d}s,
\label{eq:Efinite}\\
U_L(r)&=\boxed{S_3\left[-rG(1)+G(r)
 +\frac{m\mathrm{Wo}^2r}{2}\int_r^1\frac{W(s)}{s^2}\,\mathrm{d}s\right]}.
\label{eq:Lfinite}
\end{align}
The first term in Eq.~\eqref{eq:Efinite} enforces the transferred wall value; the integral is
the bulk Reynolds-stress contribution. This radial quadrature determines the magnitude and
direction of the leading axisymmetric component of the total drift at each finite $\mathrm{Wo}>0$.
For a single travelling wave it gives the entire leading period-averaged transport. For a mixture it gives its
axisymmetric projection. The fixed-parameter small-amplitude conditions of Sec.~\ref{sec:problem}
continue to apply.

The core Eulerian angular speed is
\begin{equation}
\Omega_m(0)=-G(1)+\frac{m\mathrm{Wo}^2}{2}\int_0^1\frac{W(s)}{s^2}\,\mathrm{d}s .
\label{eq:core}
\end{equation}
For the present radial wall, $W(1)=0$ also gives $\Omega_m'(1)=0$ in this reference-circle
mean equation.

In the viscous limit the mean azimuthal momentum equation has zero Reynolds forcing. It is $U_E''+U_E'/r-U_E/r^2=0$; regularity removes the $1/r$ solution and leaves $U_E=Kr$.
Equation~\eqref{eq:meanwall} fixes $K$, giving
\begin{align}
U_E(r)&=S_3\frac{m+2}{2}r,\label{eq:Esmall}\\
U_L(r)&=S_3\frac{m+2}{4}
\left[2r+r^{2m-3}\{m(m+1)(1-r^2)^2-2\}\right].
\label{eq:Lsmall}
\end{align}
The Eulerian contribution is a solid-body rotation, whose angular speed is fixed by the
transferred wall value $-S_3G(1)$. These are the leading coefficients as $\mathrm{Wo}\to0$.
For $m\ge2$, the latter expression is
\begin{equation}
U_L(r)=S_3\frac{m+2}{4}r
\left[2(1-r^{2m-4})+m(m+1)r^{2m-4}(1-r^2)^2\right].
\label{eq:Lpositive}
\end{equation}
The bracket is strictly positive for $0<r<1$: the first term is nonnegative and the second is
positive. Thus the total axisymmetric transport follows $S_3$ throughout the interior for
$m\ge2$. At $m=1$, Eq.~\eqref{eq:Lsmall} instead gives
$U_L=(3/2)S_3r(r^2-1)$, which keeps the opposite sign throughout the interior.
For $S_3>0$, the negative Stokes term $(3/2)S_3r(r^2-2)$ exceeds the positive Eulerian rotation
$(3/2)S_3r$ in magnitude throughout the interior.
All cases vanish at the wall. For $m\ge3$ the wall zero is simple, with
$\mathrm{d}U_L/\mathrm{d}r=-(m+2)(m-2)S_3$ at $r=1$. At $m=2$, $U_L=6S_3r(1-r^2)^2$ has a double wall zero,
so finite-Womersley-number corrections can change its sign near the wall.
The Stokes-drift reversal and the total drift are compared in
Figs.~\ref{fig:drift}(a) and~\ref{fig:transport}(a).

At $\mathrm{Wo}\to0$, the displacement per cycle is independent of the driving rate.
Writing the first-order radial wall displacement as $\mathrm{Re}\{z(\tau)e^{im\phi}\}$ gives
$z=-a e^{-i\tau}+b e^{i\tau}$. Its signed area in the complex $z$ plane is $-\pi S_3$.
A standing wave encloses zero area. This connects the viscous-limit transport with
geometric mixing by cyclic boundary motion~\cite{arrieta2015}.

For $\mathrm{Wo}\ll m\to\infty$ at fixed $\zeta=m(1-r)$, so that the viscous limit is taken first,
the coefficients satisfy
\begin{equation}
\frac{2U_S}{S_3m}\to(2\zeta^2-1)e^{-2\zeta},\qquad
\frac{2U_E}{S_3m}\to1 .
\label{eq:planar}
\end{equation}
Their sum gives the total drift on the same scale. With $k=m/R_0$ and wall amplitude
$b_w=\varepsilon R_0$, the Eulerian scale is $\omega k b_w^2/2$, the swimming speed of Taylor's
transversely waving sheet~\cite{taylor1951}. The Stokes-drift profile follows from Taylor's first-order
velocity field. His sheet is inextensible, but a planar
transverse wave needs no first-order tangential motion, so at this order it moves like the
present radial wall. We recover Tuck's waving-sheet result~\cite{tuck1968} in the planar limit $m\to\infty$
at fixed wave Reynolds number $\mathrm{Re}_k=\omega/(\nu k^2)=\mathrm{Wo}^2/m^2$.
The core angular speed of Eq.~\eqref{eq:core} then satisfies $2\Omega_m(0)/m\to(1+F)/(2F)$ with
$F=\{[1+(1+\mathrm{Re}_k^2)^{1/2}]/2\}^{1/2}$. This limit follows from Eq.~\eqref{eq:core} with the planar profile
$mf=(e^{-q\zeta}-qe^{-\zeta})/(1-q)$, $q=(1-i\mathrm{Re}_k)^{1/2}$. The Bessel profile, extrapolated in $1/m$, approaches it
within $6\times10^{-4}$ from $m=40$ and $80$ and within $1.5\times10^{-4}$ from $m=80$ and $160$, for $0.25\le\mathrm{Re}_k\le64$. The remainder
$\Omega_m(0)-(m/2)(1+F)/(2F)$, from $+1$ in the viscous limit to $-1$ at large Womersley number,
is the curvature correction; the finite-$m$ coefficients retain the circular curvature.

\emph{Tangential motion at fixed wall shape.} To isolate the effect of material motion,
we keep the radial deformation of the travelling wave and allow tangential displacement
of the wall material. A wall point with label $\phi$ is displaced by $-\varepsilon\sin(m\phi-\tau)$ in
radius, as before, and by $-\varepsilon\gamma\cos(m\phi-\tau)$ along the wall. The first wall
condition, $mf(1)=1$, is unchanged, and the second becomes $f'(1)=\gamma$, the signed ratio of
the tangential to the radial displacement amplitude. Tracking material points on the inner wall
would determine this ratio even when the travelling shape is the same. Note that Ref.~\onlinecite{liu2026} defines $\gamma$ with the
opposite sign. Every wall point still
moves periodically, so its velocity has zero mean in every fixed direction; the transferred
condition~\eqref{eq:walltransfer}, and with it Eq.~\eqref{eq:meanwall}, therefore hold as before,
and Eqs.~\eqref{eq:Efinite} and~\eqref{eq:Lfinite} apply with the profile that satisfies
$f'(1)=\gamma$. A phase difference between the two wall components would make $\gamma$ complex;
we take it real. The value $\gamma=1/m$ is the first-order inextensible wall, and for $m=1$,
$\gamma=1$ is a rigid translation. In the viscous limit $f=Ar^m+Br^{m+2}$ with
$B=(\gamma-1)/2$ and $A=1/m-B$, and
\begin{align}
U_L&=S_3\left[G(r)-rG(1)\right],\nonumber\\
G&=m\,r^{2m-3}\left[m(m-1)A^2+2m(m+1)ABr^2+(m+1)(m+2)B^2r^4\right].
\label{eq:gammaG}
\end{align}
For $m=1$ and $m=2$ this gives
\begin{align}
U_L/S_3&=\tfrac32(1-\gamma)^2\,r(r^2-1), &&m=1,\nonumber\\
U_L/S_3&=6(1-\gamma)\,r(1-r^2)\left[1-(1-\gamma)r^2\right], &&m=2.
\label{eq:gamma12}
\end{align}
For $m=1$ the first-order outline is a circle with an orbiting centre. Rigid translation
($\gamma=1$, $f=r$) gives zero total drift at every $\mathrm{Wo}$. For every other $\gamma$
the viscous-limit drift opposes the dominant wave. For $m=2$ the total drift keeps the sign of $S_3$ throughout the interior for
$0<\gamma<1$ and vanishes at $\gamma=1$, where the first-order flow is irrotational. The radial
wall, $\gamma=0$, is the marginal case with the double wall zero, which finite inertia turns into
the thin reversal of Sec.~\ref{sec:reversal}. For $\gamma<0$ the drift opposes the wave for
$r>(1-\gamma)^{-1/2}$. Near the axis the first-order inextensible wall halves the radial-wall
total drift [Fig.~\ref{fig:geometry}(b)].
For the first-order inextensible wall, set $\gamma=1/m$ and $x=r^2$ in Eq.~\eqref{eq:gammaG}. Then
\[
\frac{U_L}{S_3 r}=\frac{(m-1)(m+1)^2}{4}
\left[x^{m-2}(1-x)^2+\frac{2(1-x^m)}{m(m+1)}\right].
\]
The first term in brackets is non-negative and the second is positive for $0<x<1$.
Thus the viscous-limit total drift follows the dominant wave at every interior radius for every
integer $m\ge2$.

An independent creeping-flow calculation with tangential wall motion agrees with
Eq.~\eqref{eq:gamma12} at $m=2$ and $\gamma=\pm1/2$ (Sec.~S3).

\section{Reversals of the Stokes drift and the total drift}
\label{sec:reversal}

\begin{figure*}[tp]
\centering
\includegraphics[width=\textwidth]{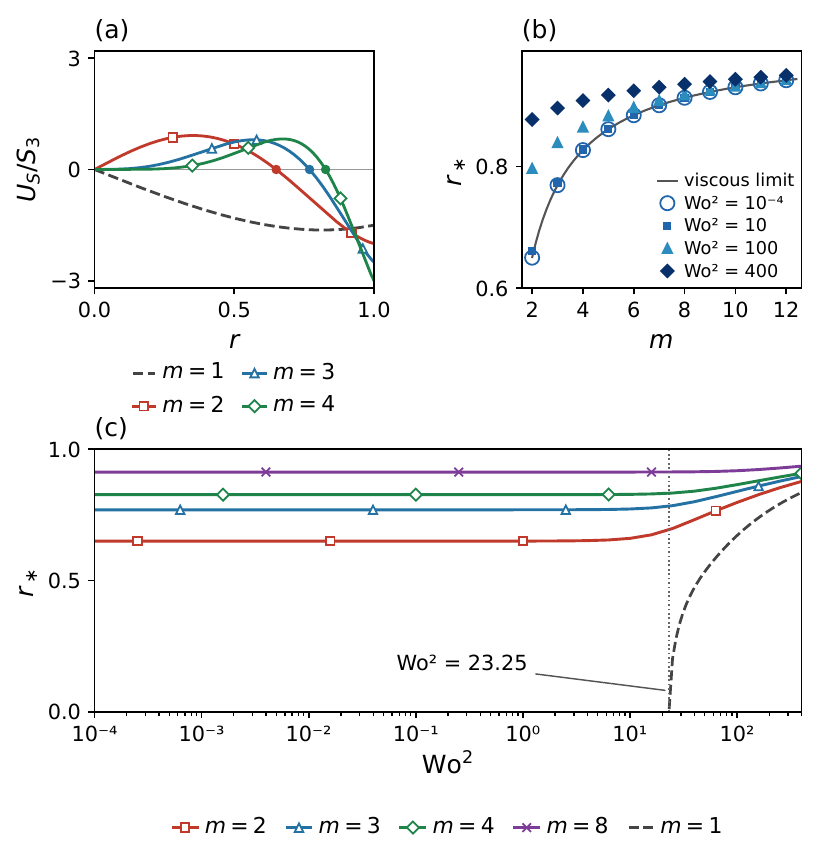}
\caption{Reversal of the ring-averaged Stokes drift for the radial wall.
(a) Viscous-limit profiles scaled by $S_3$. For $S_3>0$, positive drift follows the dominant wave
towards $+\theta$. Filled circles mark the interior zeros for $m\ge2$.
The dashed $m=1$ profile keeps one sign. (b) Reversal radius against azimuthal order.
The line gives the viscous-limit radius from Eq.~\eqref{eq:reversal}. Symbols give zeros
of the Bessel profile for each integer order from 2 to 12 at the four values of $\mathrm{Wo}^2$ in the legend.
(c) Reversal radius against $\mathrm{Wo}^2$. The dashed branch tracks the outermost detected
zero for $m=1$ from its onset near $\mathrm{Wo}^2=23.25$ (dotted line).
All profiles are leading-order results on the reference disk.}
\label{fig:drift}
\end{figure*}

Equation~\eqref{eq:elementary} vanishes inside the cavity where $m(m+1)(r^2-1)^2=2$, that is at
\begin{equation}
r_\ast^2 = 1-\sqrt{\frac{2}{m(m+1)}} ,
\label{eq:reversal}
\end{equation}
and the behaviour splits on $m$ [Fig.~\ref{fig:drift}(a)]. At $m=1$ the root sits exactly at the
origin and the axisymmetric Stokes drift keeps one sign throughout the cavity. From $m=2$ the root
moves strictly inside: for $S_3>0$ the inner Stokes component follows the wave and the outer
annulus opposes it. Equation~\eqref{eq:reversal} rises steeply with $m$ and then
flattens [Fig.~\ref{fig:drift}(b)], from $r_\ast = 0.650115$ at $m=2$ to $0.912871$ at $m=8$, so at
high $m$ the inner region occupies most of the radius and the wave-opposing annulus is thin.
The radius partitions the signs of the Stokes component. Deep in the cavity, the factor
$r^{2m-3}$ reduces its magnitude as $m$ grows, while the viscous-limit total drift keeps one
sign for every $m\ge2$ [Eq.~\eqref{eq:Lpositive}].

\emph{An interior Stokes reversal exists at every fixed Womersley number.}
Let $h=r^{-1}(|f|^2)'$, so $G=(m/4)h'$. For $m\ge2$, regularity gives $h(0)=0$,
and the radial-wall condition $f'(1)=0$ gives $h(1)=0$. Since $f(0)=0$ and $f(1)=1/m$,
$h$ cannot vanish identically. It must therefore increase and decrease inside the disk.
Thus $G$ takes both signs and
\begin{equation}
\int_0^1G(r)\,\mathrm{d}r=0,\qquad m\ge2.
\label{eq:Gintegral}
\end{equation}
This establishes existence of an interior sign change.
Equation~\eqref{eq:Gintegral} also states that the radial integral of the leading Stokes drift vanishes, so for
$m\ge2$ the radially integrated total drift equals the Eulerian volume flux through a radial cut,
$\int_0^1U_L\,\mathrm{d}r=S_3\int_0^1E_m\,\mathrm{d}r$, which is $S_3(m+2)/4$ in the
viscous limit. At $m=1$ regularity gives $h(0)=2|f'(0)|^2$, so
$\int_0^1G\,\mathrm{d}r=-|f'(0)|^2/2$, which is $-9/8$ in that limit, and
$\int_0^1U_L\,\mathrm{d}r=-3S_3/8$. The volume flux through a fixed radial cut remains the Eulerian
one, $3S_3/4$: the first-order velocity does not vanish on the axis for $m=1$, and for a single wave
the particles whose mean positions lie within the orbit radius $\varepsilon|f'(0)|$ of the axis circle
it once per period. They carry the difference between the two integrals, $S_3|f'(0)|^2/2$.
With tangential wall motion $h(1)=2\gamma/m$, so the integral in Eq.~\eqref{eq:Gintegral} is
$\gamma/2$ and an interior sign change is not guaranteed. The numerical scan below tests the
number and locations of roots.

\emph{Migration of the Stokes reversal.} Evaluating
Eq.~\eqref{eq:main} with the finite-Womersley-number Bessel profile gives one interior root
at each of the $72$ sampled pairs. The orders are $m=2,3,4,5,6,8,10,12$, and
$\mathrm{Wo}^2=10^{-4}$, $0.05$, $0.5$, $1$, $5$, $10$, $50$, $100$ and $400$.
Thus $r_\ast$ is single valued on this grid.

It moves outward
as $\mathrm{Wo}$ grows [Fig.~\ref{fig:drift}(c)]: at $m=2$ from $0.650115$ to $0.877148$, and at
$m=8$ only from $0.912871$ to $0.935263$. At large $\mathrm{Wo}$ the first-order vorticity
concentrates into a Stokes layer near the wall. The radial intensity profile changes with
$\mathrm{Wo}$, and the zeros of its derivative combination in Eq.~\eqref{eq:main}
move outward over the sampled range. High-$m$ modes are already concentrated near the wall
at small $\mathrm{Wo}$, and their reversal radii change less over this range. The scan has the perturbative scope of Sec.~\ref{sec:problem}.
The search brackets sign changes on $20001$ radii over $10^{-4}\le r\le0.995$
and refines them by Brent iteration. The single-root count applies to this interval and grid.

For the radial $m=1$ mode, a Stokes-drift zero emerges near the axis at
$\mathrm{Wo}^2\simeq23.25$. As $\mathrm{Wo}$ increases, the weak vortical tail develops
additional zeros in the core, and the dashed branch in Fig.~\ref{fig:drift}(c) tracks the
outermost detected zero. The total drift opposes the wave throughout the interior
over the sampled range $10^{-2}\le\mathrm{Wo}^2\le10^7$.

\emph{A thin total-drift reversal at small nonzero Womersley number.}
For $m=2$, expanding the first-order equation and Eq.~\eqref{eq:Lfinite} gives
$[U_L/S_3]'(1)=\mathrm{Wo}^4/640+O(\mathrm{Wo}^8)$.
With $d=1-r$, the near-wall balance is
\begin{equation}
U_L/S_3=24d^2-\frac{\mathrm{Wo}^4d}{640}
 +O(d^3+\mathrm{Wo}^4d^2+\mathrm{Wo}^8d).
\label{eq:smallqwall}
\end{equation}
Consequently, for sufficiently small nonzero $\mathrm{Wo}$ a nonwall zero satisfies
$1-r=\mathrm{Wo}^4/15360+O(\mathrm{Wo}^8)$. This is a reversal of the reference-disk coefficient in a
thin annulus; its finite-amplitude interpretation requires the wall-swept-strip
qualification in Sec.~\ref{sec:functional} and, at fixed physical points, a wall displacement
small compared with the annulus width $\mathrm{Wo}^4/15360$, a condition much stronger than
$\varepsilon\mathrm{Wo}\ll1$. For $m=2$ and $0\le\gamma\le1/2$, tangential wall motion adds the
viscous-limit wall slope $12\gamma(\gamma-1)$ of Eq.~\eqref{eq:gamma12}. The near-radial annulus exists
below a threshold in $\gamma$, which is $\mathrm{Wo}^4/7680$ at leading order and $0.012$ at $\mathrm{Wo}^2=10$.

The wall slope at any fixed Womersley number follows from
Eqs.~\eqref{eq:bc} and~\eqref{eq:Lfinite}:
\begin{equation}
\left[\frac{U_L}{S_3}\right]'(1)
=(m-4)G(1)-\frac{m(m+2)}{2}.
\label{eq:wallslope}
\end{equation}
For $m=4$ it is exactly $-12$. For $m=3$ it changes sign at
$\mathrm{Wo}^2\simeq404.463$, with a positive slope corresponding to negative
$U_L/S_3$ immediately inside the wall. The expansion of $G(1)$ in
Appendix~\ref{app:core} gives the large-Womersley-number slope
$(4-m)\mathrm{Wo}/(2\sqrt2)+O(1)$ at fixed $m$.
This wall-local criterion does not determine the sign throughout the interior.

\begin{figure*}[tp]
\centering
\includegraphics[width=\textwidth]{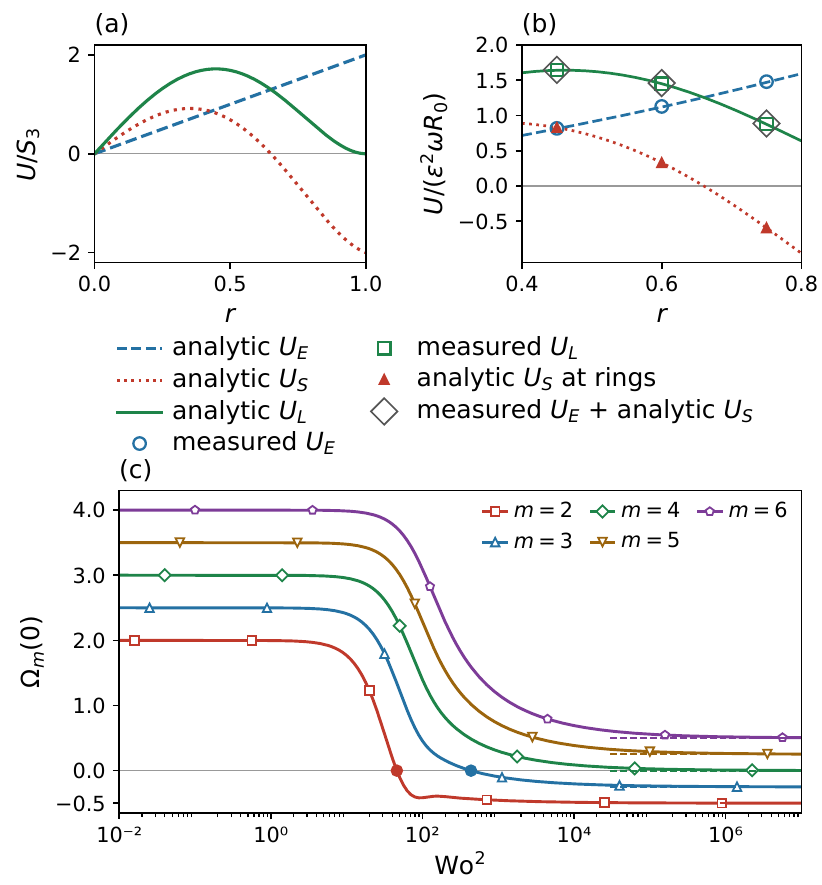}
\caption{Ring-averaged transport and core rotation for a positive travelling wave.
(a) Coefficients of the Eulerian mean, Stokes drift and total drift for $m=2$ in the viscous limit,
normalized by $S_3$. (b) At $\mathrm{Wo}^2=10$ for the same mode, curves give
the three leading coefficients. Symbols give the Eulerian mean and the total drift measured from tracers in the
moving-domain computation, the analytic Stokes drift, and the sum of the measured Eulerian
mean and analytic Stokes drift. The base run uses $\varepsilon=0.02$, 96 steps per period and
physical-ring release. Velocities are scaled by $\varepsilon^2\omega R_0$.
(c) Core Eulerian angular speed $\Omega_m(0)$ against $\mathrm{Wo}^2$ for modes 2 to 6.
Dashed lines mark the large-$\mathrm{Wo}$ limits $(m-4)/4$. Filled circles mark the sign changes
at $\mathrm{Wo}^2\simeq46.0$ for $m=2$ and $437.4$ for $m=3$.
For $m\ge3$ the core angular speed also gives the near-axis limit of $U_L/(S_3r)$.}
\label{fig:transport}
\end{figure*}

\emph{Directions of the Eulerian mean and total drift at finite $\mathrm{Wo}$.}
At $m=2$, $\mathrm{Wo}^2=100$ and $r=0.60$, the wall and bulk terms in
Eq.~\eqref{eq:Efinite} are $2.366$ and $-2.572$, giving $E_m=-0.206$.
At this sample the two terms nearly cancel, so the transferred wall value alone does not set the core flow.
At $r=0.60$, $E_m$ changes sign at $\mathrm{Wo}^2\simeq71.7184$;
the core angular speed of Eq.~\eqref{eq:core} changes sign at $\mathrm{Wo}^2\simeq46.0077$.

For $m=2$ and $\mathrm{Wo}^2=100$, independent refinement of the radial boundary-value problem
and high-precision quadrature place the total-drift zero at $r=0.968452$.
In the outer annulus the Eulerian and Stokes contributions nearly cancel. Even at its
largest, the reversed speed reaches only 2.7 per cent of the positive peak (Sec.~S5).

At $\mathrm{Wo}^2=10$, the value used in the moving-domain test, the zero lies closer to the wall
at $r=0.994542$. Beyond it the total drift coefficient reaches only $-2.0\times10^{-4}$ against a
peak of 1.64. For $\varepsilon=0.02$ the minimum wall radius is $0.979796$. The wall thus
sweeps across the zero at $\mathrm{Wo}^2=10$, while the zero at $\mathrm{Wo}^2=100$ remains in the fluid
throughout the cycle. Resolving either reversal in particle trajectories requires a wall
displacement and spatial resolution small compared with the width of the reversed annulus.

For $m=3$, Eq.~\eqref{eq:core} changes sign near $\mathrm{Wo}^2=437.382$.
Since $G=O(r^3)$ near the axis, a negative core Eulerian angular speed also reverses the total drift
sufficiently close to the axis. For example, $U_L/S_3=-0.006137$ at
$\mathrm{Wo}^2=1600$, $r=0.05$.

At large Womersley number the transferred wall value grows as
$-G(1)=\mathrm{Wo}/(2\sqrt2)+1/4+O(\mathrm{Wo}^{-1})$, and the Reynolds-stress integral in
Eq.~\eqref{eq:core} cancels its $O(\mathrm{Wo})$ part. A Stokes-layer expansion of the Bessel profile
(Appendix~\ref{app:core}) gives
\begin{equation}
\Omega_m(0)=\frac{m-4}{4}+O(\mathrm{Wo}^{-1}),
\label{eq:corelimit}
\end{equation}
and the exact profile at $\mathrm{Wo}=160$, $640$ and $2560$ agrees for $m=1,2,3,4,6$, with departures that
fall as $1/\mathrm{Wo}$. Since $G=O(r^{2m-3})$ near the axis for $m\ge3$, among $m\ge2$ the near-axis total
reverses at large Womersley number only for $m=3$; at $m=2$ it tends to $S_3r/2$.
For $m=4$, the next term is $\Omega_4(0)=6\sqrt2/\mathrm{Wo}+O(\mathrm{Wo}^{-2})$,
so the core approaches zero from above.
At fixed interior radius $0<r<1$, the potential part $f\to r^m/m$ gives
\begin{equation}
\frac{U_L}{S_3}\longrightarrow
r\left[\frac{m-4}{4}+(m-1)r^{2m-4}\right],\qquad m\ge2.
\label{eq:interiorlimit}
\end{equation}
For a single radial travelling wave, Eq.~\eqref{eq:interiorlimit} recovers the outer profile of
Ilin and Morgulis~\cite{ilin2019} for a sinusoidal helical wave as its axial wavenumber tends
to zero [their Eqs.~(19)--(26)]. They take the wall amplitude comparable to the Stokes-layer
thickness. Here we first take the small-amplitude limit at fixed $\mathrm{Wo}$.
For $m=3$ the limiting reversed region extends to $r=2^{-3/2}\simeq0.354$.
At $\mathrm{Wo}^2=1600$ its outer zero is still at $r=0.240192$.

Over $10^{-2}\le\mathrm{Wo}^2\le10^7$, only modes 2 and 3 among the five plotted change
the sign of their core Eulerian rotation [Fig.~\ref{fig:transport}(c)]. The $m=2$ curve has
a shallow minimum of $-0.418$ near $\mathrm{Wo}^2=93$ and a local maximum of $-0.391$ near
$\mathrm{Wo}^2=157$.

Between the wall and the interior limit~\eqref{eq:interiorlimit} lies the oscillatory boundary
layer of a progressive wave. At fixed $\eta=(1-r)\mathrm{Wo}/\sqrt2$ the expansion of
Appendix~\ref{app:core} gives the total drift in the layer in closed form,
\begin{align}
\frac{U_L}{S_3}={}&\frac m4\left[5-8e^{-\eta}\cos\eta+3e^{-2\eta}\right]\nonumber\\
&-2\left(1-e^{-\eta}\cos\eta\right)+O(\mathrm{Wo}^{-1}).
\label{eq:layer}
\end{align}
The part proportional to $m$ is the mass-transport profile of
Longuet-Higgins~\cite{longuethiggins1953} for slip-velocity amplitude $\varepsilon\omega R_0$ and
phase speed $\omega R_0/m$, so that $U^2/c=m$ on the present scale. The part independent of $m$ is a
curvature term. It falls from zero at the wall, with slope $-2$ in $\eta$, to $-2.13$ at $\eta=3\pi/4$
and returns to $-2$ at the layer edge, where $U_L/S_3=(5m-8)/4$: the planar $5/4$ law reduced by
curvature. The exact Bessel profile approaches Eq.~\eqref{eq:layer} as $1/\mathrm{Wo}$: at
$\mathrm{Wo}=10^5$ the two differ by less than $4\times10^{-4}$ for $\eta\le6$ at $m=1$, $2$ and $3$.

Near the wall the leading slope is $m/2-2$. For $m=2$ the outer profile $U_L/S_3=r/2$
follows the wave, while curvature reverses the total drift just inside the wall.
The near-wall drift also reverses for $m=3$, as Eq.~\eqref{eq:wallslope} predicts. For higher modes the leading
profile in Eq.~\eqref{eq:layer} can be written as
\[
\frac{m-4}{4}\left[5-8e^{-\eta}\cos\eta+3e^{-2\eta}\right]
+3\left|1-e^{-(1-i)\eta}\right|^2 .
\]
The bracket is non-negative, so the total drift in the layer follows the wave or vanishes for $m\ge4$.
For $m=1$ the profile reduces to $-(3/4)(1-e^{-2\eta})$ and opposes the wave throughout the layer.

For the same travelling shape, the leading inward slope of the large-$\mathrm{Wo}$ total drift
becomes $\tfrac12(1-\gamma)(m-4+3m\gamma)$ when the wall also moves tangentially.
For $m\ge2$ the drift just inside the wall opposes the wave when $\gamma<(4-m)/(3m)$
or $\gamma>1$. At either equality the linear term vanishes.
The full layer profile for $f'(1)=\gamma$ is
\begin{align}
\frac{U_L}{S_3}={}&2(1-\gamma)(m-1)\left(1-e^{-\eta}\cos\eta\right)\nonumber\\
&-\frac{3m}{4}(1-\gamma)^2\left(1-e^{-2\eta}\right)+O(\mathrm{Wo}^{-1}).
\label{eq:layergamma}
\end{align}
The corresponding core limit is $(m-1)(1-2\gamma)-(3m/4)(1-\gamma)^2$.
A reversed annulus has wave-following drift farther inside. On the branch containing the radial wall for $m=2,3$, the large-$\mathrm{Wo}$
threshold is $1/3$ and $1/9$, respectively. For $m=2$ this threshold in $\gamma$ rises from
$\mathrm{Wo}^4/7680$ at small $\mathrm{Wo}$ through $0.16$ at $\mathrm{Wo}^2=100$.
The first-order inextensible wall, $\gamma=1/m$, has a positive leading inward slope for $m\ge2$.
The quadrature gives wave-following drift immediately inside that wall for $m=2$ to $6$
over $1\le\mathrm{Wo}^2\le10^8$. Its core limit is $(m-1)(m-5)/(4m)$,
which is negative for $m=2$, $3$ and $4$. The near-axis total drift therefore reverses for $m=3$ and $m=4$,
beyond the core sign changes at $\mathrm{Wo}^2\simeq146$ and $1330$; for $m=2$ the core Eulerian
rotation changes sign at $\mathrm{Wo}^2\simeq38.1$, while the particles near the axis follow the wave.

Since $\omega R_0/c_0=\mathrm{Wo}^2\nu/(c_0R_0)$, a compact cavity requires
$\mathrm{Wo}^2\ll c_0R_0/\nu$, which is $1.5\times10^5$ for water at $R_0=100\,\mu\mathrm{m}$ and
$1.5\times10^7$ at $R_0=1\,\mathrm{cm}$; the limits~\eqref{eq:corelimit}
and~\eqref{eq:interiorlimit}, approached only for $\mathrm{Wo}^2$ of order $10^5$ to $10^6$,
therefore describe centimetre-scale cavities.

\begin{figure*}[t]
\centering
\includegraphics[width=\textwidth]{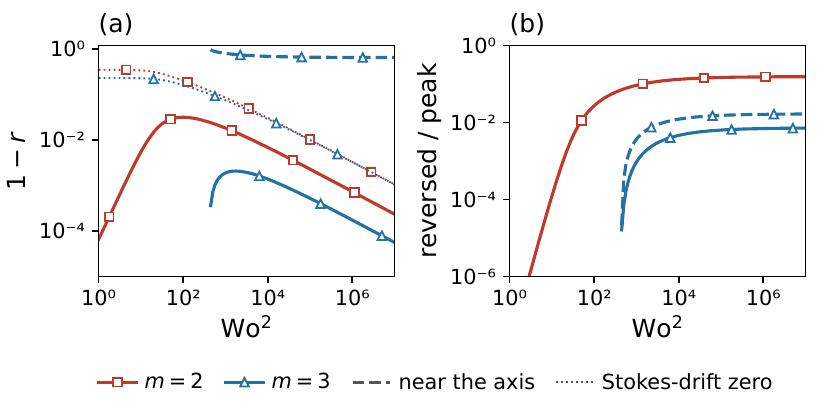}
\caption{Reversals of the total drift, radial wall. (a) Distance from the wall, $1-r$, of the zeros
of the leading coefficient $U_L/S_3$ against $\mathrm{Wo}^2$ for $m=2$ (open squares) and $m=3$ (open triangles): the
near-wall zero (solid), the outer edge of the near-axis reversed region for $m=3$ (dashed) and, for
comparison, the zero $r_\ast$ of the Stokes component (dotted). (b) The magnitude of the most negative $U_L/S_3$ in
each reversed region divided by the largest $U_L/S_3$ in the cavity at the same $\mathrm{Wo}^2$;
line styles as in (a). All quantities are leading-order coefficients on the reference disk.}
\label{fig:reversal}
\end{figure*}

Over $1\le\mathrm{Wo}^2\le10^7$ the reversed annulus for the radial wall at $m=2$ is widest
near $\mathrm{Wo}^2=100$, where it spans 3.2 per cent of the cavity radius [Fig.~\ref{fig:reversal}(a)].
As $\mathrm{Wo}$ increases further, the annulus thins with the Stokes layer. Its most negative drift approaches
the minimum $-0.100$ of Eq.~\eqref{eq:layer}, or 15.5 per cent of the peak $0.649$ in magnitude
[Fig.~\ref{fig:reversal}(b)]. For $m=3$ the near-wall
reversal tends to 0.72 per cent of the peak and the near-axis
reversal to 1.67 per cent; the latter is $-0.034$ at $r=24^{-1/2}\simeq0.204$, the minimum
of Eq.~\eqref{eq:interiorlimit}. Each reversal is narrow or weak beside a drift that follows the
wave over most of the cavity.

\section{Verification against a moving-domain computation}
\label{sec:verification}

The primitive equations are solved on the deforming domain by an arbitrary
Lagrangian-Eulerian method. The base comparison uses $m=2$, $\mathrm{Wo}^2=10$,
$\varepsilon=0.02$ and rings at $r=0.45$, $0.60$ and $0.75$.
We measure the ring-averaged total drift $U_L$ from tracer displacements and the
Eulerian mean $U_E$ at fixed physical points over the same periods and rings.
We compare these measurements with the leading coefficients and use the analytic Stokes drift $U_S$
to reconstruct the total drift from the measured Eulerian mean.

The solver extends the moving-domain method used for the standing cavity~\cite{liu2026}
to the travelling and mixed wall motions. It evolves vorticity and streamfunction on a radial map
onto the area-preserving moving wall that is the identity for $\rho\le0.1$, with finite differences
in radius, Fourier collocation in angle, a Crank--Nicolson viscous step and explicit nonlinear terms.
The base mesh has $96$ radial nodes and $24$ angular collocation points, with $96$ steps per period;
the computation starts from zero vorticity, discards three periods and measures the next three.
Sixteen tracers per ring are released on the physical ring of radius $r$, the base observable, and the
Eulerian mean uses sixteen fixed physical sampling points per ring. A second observable, the mapped
release, starts each tracer at the first-order displaced position of a reference point on the ring,
so that the tracers' mean radius equals $r$ to within $2.7\times10^{-3}\,\varepsilon^2$; on the physical
ring it exceeds $r$ by $1.71\varepsilon^2$ at $r=0.60$. The supplementary material gives the protocol and
its solver tolerances (Sec.~S1) and varies the solver step, the tracer integration, the grid, the
startup and the averaging window separately (Sec.~S2).

\begin{table}[t]
\caption{Moving-domain departures from the leading coefficients for the radial wall at $m=2$ and $\mathrm{Wo}^2=10$. Here
$\delta_E=U_E/U_E^{(0)}-1$ and $\delta_L=U_L/U_L^{(0)}-1$, with $U^{(0)}$ the leading coefficients of
Eqs.~\eqref{eq:Efinite} and \eqref{eq:Lfinite}. Physical ring: tracers released on the ring of radius $r$ (the base
observable); mapped release: tracers released at the first-order displaced reference points. The numerical part is
extrapolated to zero solver step from $96$ to $768$ steps per period. The coefficients $c$ multiply $\varepsilon^2$ in
$\delta$ and come from fits over $\varepsilon=0.01$ to $0.04$ on $96$ radial nodes at $384$ steps per period. Tracer
rows other than the base row use four Heun substeps with quadratic interpolation in time.}
\label{tab:budget}
\begin{ruledtabular}
\begin{tabular}{lccc}
 & $r=0.45$ & $r=0.60$ & $r=0.75$\\
\hline
Base, $\varepsilon=0.02$: $100\delta_E$ & $+0.71$ & $+0.79$ & $+0.75$\\
\quad $100\delta_L$, physical ring & $+0.27$ & $+0.11$ & $+0.44$\\
Numerical, $\Delta\tau\to0$: $100\delta_E$ & $-0.07$ & $-0.02$ & $+0.01$\\
\quad $100\delta_L$, physical ring & $-0.02$ & $-0.03$ & $-0.10$\\
\quad $100\delta_L$, mapped release & $-0.02$ & $-0.03$ & $-0.08$\\
$c_E$, fixed points & $+2.64$ & $+2.60$ & $+2.45$\\
$c_L$, physical ring & $-2.0$ & $-10.3$ & $-12.2$\\
$c_L$, mapped release & $+1.32$ & $-0.91$ & $-0.17$\\
\end{tabular}
\end{ruledtabular}
\end{table}

On all three physical rings the base computation gives positive tracer drift for $S_3>0$,
including at $r=0.75$, where the Eulerian mean and the Stokes drift oppose each other
[Fig.~\ref{fig:transport}(b)]. The base Eulerian and tracer values exceed their leading
predictions by less than one per cent (Table~\ref{tab:budget}).

Most of this departure comes from the solver step. Three successive halvings give convergence
orders between 0.93 and 0.96, consistent with the treatment of the nonlinear terms at the old time level.
Varying the amplitude at fixed resolution separates the numerical departure from the part
proportional to $\varepsilon^2$. After extrapolation to zero solver step and correction for the
radial grid, the numerical residual falls below the time-extrapolation uncertainty (Sec.~S3).
The moving-domain computation reproduces the leading Eulerian and total coefficients
on the three rings to within 0.11 and 0.17 per cent when the absolute remaining numerical
part and this uncertainty are added.
At these rings the bulk Reynolds-stress term in Eq.~\eqref{eq:Efinite} opposes the wall
term. Its magnitude ranges from about 6 to 15 per cent of the radial-wall Eulerian mean.

At $r=0.60$ the fitted tracer amplitude coefficient $c_L$ changes from $-10.3$ for
physical-ring release to $-0.91$ for mapped release (Table~\ref{tab:budget}).
This dependence on release position persists in creeping flow (Sec.~S3).
Coarser runs also test the approach to the viscous limit and the dependence on wave composition
(Sec.~S4). The standing-wave Eulerian ring mean vanishes to roundoff, and at fixed
intensity the total drift measured for the mixed wave scales with $S_3$ to within the accuracy of
the travelling-wave control.

The moving-domain computation also tests the predicted change in drift at fixed wave shape.
The wall material now slides along that shape, with the point labelled $\phi$ at angle
$\phi-\varepsilon\gamma\cos(m\phi-\tau)$ (Sec.~S6). At $m=2$ and $\mathrm{Wo}^2=10$,
tangential wall motion lowers the predicted total drift coefficient at $r=0.60$ from
$1.45$ for the radial wall to $0.933$ for the first-order inextensible wall ($\gamma=1/2$).
The moving-domain computation gives $0.937$ for the inextensible wall.
After amplitude and solver-step extrapolation, the moving-domain results agree with the
leading Eulerian and total drift coefficients to within 0.20 and 0.13 per cent, respectively.
Section~S6 and supplementary Table~S2 give the error bounds for each ring. Here the opposing bulk term has magnitude 7.4 to 18.4 per cent of the Eulerian mean.

A second implementation of the reference-disk expansion, a curved P3/P2 finite-element code,
computes the first-order field, transfers the second-order boundary condition and constructs the
Stokes drift and the total drift for $m=2$, $3$ at $\mathrm{Wo}^2=10^{-4}$, $1$, $10$ and $100$.
For the radial wall at $m=2$ and $\mathrm{Wo}^2=100$, the finest mesh locates the near-wall
zero of the total drift within $2.5\times10^{-6}$ of the quadrature prediction, $r\simeq0.96845$.
Across the eight cases the largest error in the Stokes-drift zero falls from $2.4\times10^{-6}$
to $6.1\times10^{-8}$ under refinement (Sec.~S5).

\emph{Use of AI tools.} Claude Opus 5/5.5 and Fable 5.2 (Anthropic, through Claude Code)
and GPT 6/6.1/6.2 (OpenAI, through Codex) assisted with literature verification, code development,
numerical execution and data analysis under the authors' direction to cross-check the
derivations and their numerical implementation. Deterministic numerical solvers produced the results, and plotting scripts produced the figures.

\section{Discussion}
\label{sec:discussion}

Exchanging the intensities of the two travelling waves reverses the leading
axisymmetric total drift. For a radial wall and $m\ge2$, particles follow the stronger wave
throughout the interior in the viscous limit, although the Stokes drift changes sign with
radius for all $\mathrm{Wo}$. At $m=2$, $\mathrm{Wo}^2=10$ and $r=0.75$, tracers still
follow the wave where the Stokes drift opposes it.

For the radial wall at $m=2$, the double wall zero in the viscous limit opens into a
thin reversed annulus at sufficiently small nonzero $\mathrm{Wo}$. At $m=3$ the total drift
can reverse both near the wall and near the axis. In the radial-wall profiles examined here,
the reversed speed stays below a sixth of the peak. The strongest reversal lies in the
Stokes layer, where curvature reduces the planar Longuet-Higgins profile.

At fixed travelling shape and large $\mathrm{Wo}$, particles just inside the wall oppose
the wave for $\gamma<(4-m)/(3m)$ or $\gamma>1$ ($m\ge2$). Preserving arc length to
first order restores wave-following drift there, while the near-axis total drift reverses for $m=3$ and $4$.
For $m=2$ the core Eulerian rotation reverses at $\mathrm{Wo}^2\simeq46$ for the
radial wall and at $\mathrm{Wo}^2\simeq38.1$ for the first-order inextensible wall.
Particles near the axis still follow the wave. Distinguishing fixed-point velocity from
particle displacement also matters for reflux in peristaltic pumping~\cite{shapirojaffrin1971}.
Bouvard et al. reduced stroboscopic sampling bias by averaging the measured velocity field
with a copy rotated through $\pi$~\cite{bouvard2017}.

At $m=2$ and $\mathrm{Wo}^2=10$, moving-domain trajectories recover the leading total
drift for both walls within 0.2 per cent on three interior rings where particles follow the wave.
At finite amplitude the drift depends on tracer release and endpoint estimation even in creeping flow.
Resolving the predicted near-wall reversal in trajectories requires both the wall displacement
and the grid spacing to be small compared with the width of the reversed annulus.

\section*{Supplementary Material}
The supplement gives the moving-domain protocol and sensitivities (Secs.~S1--S2),
amplitude extrapolation and creeping-flow benchmarks (Sec.~S3), wave-composition and small-$\mathrm{Wo}$
controls (Sec.~S4), finite-element verification (Sec.~S5) and tests of tangential wall motion
(Sec.~S6 and Table~S2).

\begin{acknowledgments}
J.O. was supported by a Chancellor's Undergraduate Research Award (CURA) from the University of
Illinois Chicago.
Claude Opus 5/5.5 and Fable 5.2 (Anthropic, through Claude Code) and GPT 6/6.1/6.2 (OpenAI, through
Codex) provided editorial assistance with the manuscript under the authors' direction; their use in
the research is described in Sec.~\ref{sec:verification}.
Kimi K3.1 drew our attention to earlier work on steady streaming between vibrating walls.
Gemini 4 Carbon and Neon provided readability improvement suggestions. The authors are responsible for the
accuracy, originality and final content of this work.
\end{acknowledgments}

\section*{Author Declarations}
\subsection*{Conflicts of Interest}
The authors state no conflict of interest.

\subsection*{Author Contributions}
\textbf{Zijian Liu}: Conceptualization (equal); Formal analysis (equal); Investigation (equal);
Methodology (equal); Validation (equal);
Writing -- original draft (equal); Writing -- review \& editing (equal).

\textbf{Julian Olszewski}: Conceptualization (equal); Formal analysis (equal);
Funding acquisition (equal); Investigation (equal); Methodology (equal);
Validation (equal); Writing -- original draft (equal);
Writing -- review \& editing (equal).

\textbf{Jie Xu}: Conceptualization (lead); Formal analysis (lead); Project administration (lead); Resources (lead);
Supervision (lead); Writing -- review \& editing (equal).

\section*{Data Availability}
The data supporting this study will be publicly available through the University of Illinois Chicago's
INDIGO repository upon acceptance of the manuscript.

\appendix
\section{Core rotation and total drift in the Stokes layer at large Womersley number}
\label{app:core}
Let $\alpha=\mathrm{Wo}/\sqrt2$, $\Lambda=\alpha(1-i)$ and $x=1-r$.
For fixed $m$, in the Stokes layer $x=O(\alpha^{-1})$, the Bessel part of $f$,
normalized to unity at the wall, is
$g=r^{-1/2}e^{-\Lambda x}[1+O(\alpha^{-2})]$.
Thus $\sigma=g'(1)=\Lambda-1/2+O(\alpha^{-1})$, and Bessel's equation gives
$g''(1)=\Lambda^2-\sigma+m^2$.
Equation~\eqref{eq:bc} gives $f=Cr^m+\tilde B_m g$, with
$C=\sigma/[m(\sigma-m)]$ and $\tilde B_m=-mC/\sigma$. Their expansions are
\begin{align}
C&=\frac1m+\frac1\Lambda+\frac{m+1/2}{\Lambda^2}+O(\alpha^{-3}),\\
\tilde B_m&=-\frac1\Lambda-\frac{m+1/2}{\Lambda^2}+O(\alpha^{-3}).
\end{align}
It follows that
$f''(1)=(m\sigma-\Lambda^2-m^2)/(\sigma-m)=-\Lambda-1/2+O(\alpha^{-1})$
and $-G(1)=\alpha/2+1/4+O(\alpha^{-1})$.
The potential term $Cr^m$ has constant phase. The remaining terms in
$W=\mathrm{Im}(\bar f f')$ are
\begin{align}
W=\mathrm{Im}\big[&\bar C\tilde B_m r^m g'
 +\overline{\tilde B_m}C m r^{m-1}\bar g\nonumber\\
 &+|\tilde B_m|^2\bar g g'\big].
\label{eq:Wcross}
\end{align}
They are confined to the layer, where, to the order needed for the constant core limit,
\begin{align}
W={}&-\frac1m\left[1+\frac m\alpha-\left(m-\tfrac12\right)x\right]
 e^{-\alpha x}\sin\alpha x\nonumber\\
&+\frac{e^{-\alpha x}}{2\alpha}(\sin\alpha x+\cos\alpha x)
 -\frac{e^{-2\alpha x}}{2\alpha}+O(\alpha^{-2}).
\end{align}
The phases of $C$ and $\tilde B_m$ combine through
$m/\Lambda+m/\bar\Lambda=m/\alpha$.
With $s^{-2}=1+2x+O(x^2)$,
\begin{equation}
\int_0^1\frac{W(s)}{s^2}\,\mathrm{d}s
=-\frac1{2\alpha m}+\frac{m-5}{4m\alpha^2}+O(\alpha^{-3}).
\end{equation}
Multiplying by $m\mathrm{Wo}^2/2=m\alpha^2$ and adding $-G(1)$ gives
Eq.~\eqref{eq:corelimit}.

At $m=4$ that constant limit vanishes, so one further order determines the sign.
Writing $y=\alpha x$ and $\ell=1-i$, the normalized Bessel profile is
\begin{equation}
g=e^{-\ell y}\left[1+\frac{y}{2\alpha}
 +\frac1{\alpha^2}\left(\frac{3y^2}{8}-\frac{63y}{8\ell}\right)
 +O(\alpha^{-3})\right].
\end{equation}
Substitution into Eq.~\eqref{eq:Wcross} and the wall condition gives
\begin{align}
\int_0^1\frac{W(s)}{s^2}\,\mathrm{d}s
 &=-\frac1{8\alpha}-\frac1{16\alpha^2}+\frac{93}{128\alpha^3}
 +O(\alpha^{-4}),\\
-G(1)&=\frac\alpha2+\frac14+\frac{99}{32\alpha}+O(\alpha^{-2}).
\end{align}
Consequently $\Omega_4(0)=6/\alpha+O(\alpha^{-2})$
$=6\sqrt2/\mathrm{Wo}+O(\mathrm{Wo}^{-2})$, which is positive for sufficiently large
Womersley number.

\emph{Total drift in the Stokes layer.} Let the second wall condition be $f'(1)=\gamma$ and write
$\beta=1-\gamma$, $\eta=\alpha x$, $c=e^{-\eta}\cos\eta$ and $s=e^{-\eta}\sin\eta$. The coefficients
become $\tilde B_m=-\beta/(\sigma-m)$ and $C=1/m-\tilde B_m$. At fixed $\eta$ the profile is
$f=1/m+[\beta(1-e^{-\ell\eta})/\ell-\eta]/\alpha+O(\alpha^{-2})$, and one further order gives
\begin{align}
G={}&-\frac{\alpha\beta}{2}(c+s)+(m-1)-\frac{3\beta}{4}(2m-1)c\nonumber\\
&-\frac{m\beta^2}{2}\left(s-e^{-2\eta}\right)+\frac{\beta}{4}(2m-3)\eta(c+s)+O(\alpha^{-1}),\\
E_m={}&\frac{\alpha\beta}{2}(c+s)-g_0+\frac{\beta}{4}(2m-5)(1-c)\nonumber\\
&+\frac{m\beta^2}{4}\left(2s-1+e^{-2\eta}\right)-\frac{\beta}{4}(2m-3)\eta(c+s)+O(\alpha^{-1}),
\end{align}
with $g_0=(m-1)-3\beta(2m-1)/4+m\beta^2/2$, so that $G(1)=-\alpha\beta/2+g_0+O(\alpha^{-1})$.
The $O(\alpha)$ terms cancel in the sum, as do the terms in $s$ and $\eta(c+s)$, which leaves
Eq.~\eqref{eq:layergamma}; $\gamma=0$ gives Eq.~\eqref{eq:layer}. At the layer edge the total drift coefficient $U_L/S_3$ is
$2\beta(m-1)-3m\beta^2/4$. The potential part contributes $m-1$ to the Stokes drift coefficient $G$ there, and
the difference is the core limit $(m-1)(1-2\gamma)-(3m/4)(1-\gamma)^2$, which is
Eq.~\eqref{eq:corelimit} at $\gamma=0$.

\end{document}

\renewcommand{\thesection}{S\arabic{section}}
\renewcommand{\theequation}{S\arabic{equation}}
\renewcommand{\thetable}{S\arabic{table}}
\renewcommand{\thefigure}{S\arabic{figure}}

\title{Supplementary material for ``Lagrangian transport driven by travelling waves in a radially deforming circular cavity''}

\author{Zijian~Liu}
\thanks{Z. Liu and J. Olszewski contributed equally to this work.}
\affiliation{Department of Mechanical and Industrial Engineering,
University of Illinois Chicago, Chicago, Illinois 60607, USA}

\author{Julian~Olszewski}
\affiliation{Department of Mechanical and Industrial Engineering,
University of Illinois Chicago, Chicago, Illinois 60607, USA}

\author{Jie~Xu}
\email[Author to whom correspondence should be addressed: ]{jiexu@uic.edu}
\affiliation{Department of Mechanical and Industrial Engineering,
University of Illinois Chicago, Chicago, Illinois 60607, USA}

\date[]{}

\maketitle

This supplement details the moving-domain verification of Sec.~\ref{M-sec:verification}.
Numbers without the prefix S refer to the main text.

\section{Moving-domain protocol}
\label{sec:smprotocol}
The solver evolves vorticity and streamfunction on a blended
radial map $r=\rho\,[1+\beta(\rho)(R-1)]$ with instantaneous dimensionless wall radius $R$.
The blend satisfies $\beta=0$ for $\rho\le0.1$, so that the map is the identity in the core and carries the wall displacement near
the boundary. The enclosed area is preserved exactly; the interior map is not locally area
preserving. For $0.1<\rho<1$, the blend is
$\beta=\eta(u)/[\eta(u)+\eta(1-u)]$, with $u=(\rho-0.1)/0.9$,
$\eta(u)=e^{-1/u}$ for $u>0$ and $\eta(u)=0$ otherwise.
The uniform radial grid has nodes $\rho_i=i/N$, $i=1,\ldots,N$.
Radial derivatives use fourth-order five-point interior stencils and second-order
stencils in the two rows nearest each end; angular derivatives use Fourier collocation.
The identity map near the axis permits the usual regular radial behaviour of each
Fourier harmonic to be imposed in the harmonic solves.
The solver uses a Crank--Nicolson viscous step, explicit nonlinear terms, harmonic Poisson solves,
and a real-basis influence iteration to close the wall vorticity.
The nonlinear terms are evaluated at the old time level; the combined update therefore
does not inherit second-order accuracy from its viscous step alone.
The wall-vorticity boundary residual is below $10^{-11}$ at every accepted base-run step.
The Poisson and viscous solves iterate a splitting until the relative iterate change falls
below $10^{-12}$, with a cap of $200$ iterations. In all 16 runs of Sec.~\ref{sec:smsens}
every one of the 521\,110 inner solves met this tolerance in at most 30 iterations,
and none reached the cap; the interior discrete Poisson residual stayed below
$4.3\times10^{-10}$. A repeat of the base protocol with the tolerance tightened to $10^{-13}$, failing
on any capped solve, changes the ring means by at most $7.6\times10^{-12}$ relative; at $10^{-14}$ the
iteration stalls near a relative change of $5\times10^{-14}$, its round-off floor.

The base mesh has $96$ radial nodes and $24$ angular collocation points, with solver and
tracer steps $\Delta\tau=2\pi/96$. It starts from zero vorticity, discards three periods,
and measures the next three. Sixteen equally spaced tracers per ring advance by Heun
integration using velocities from the two endpoint snapshots of each step.
The measured total drift $U_L$ equals the initial ring radius times the mean unwrapped
angular displacement per unit time. The Eulerian mean uses sixteen fixed
physical sampling points per ring over the same window.
Coarse controls use $64$ radial nodes, $16$ angular points and $64$ steps per period,
discarding two periods and averaging two for the pure-wave amplitude controls
at $\varepsilon=0.04$, $0.02$ and $0.01$. The mixed-wave comparison instead discards
three periods and averages three for both its mixed and matched travelling runs.
The slowest linear mean-swirl mode has e-folding time
$\mathrm{Wo}^2/j_{1,1}^2$ in units of $1/\omega$, with $j_{1,1}^2\simeq14.682$. Here
$\mathrm{Wo}^2=\omega R_0^2/\nu$ for angular frequency $\omega$, cavity radius $R_0$
and kinematic viscosity $\nu$.
Twenty-five such e-foldings take about $0.271\mathrm{Wo}^2$ periods,
which gives a startup scale for changing Womersley number.

\section{Separated numerical sensitivities}
\label{sec:smsens}
Each run evolves one field and advances several tracer sets in it, so that the solver step, the tracer
integration, the grid and the tracer observable can be varied independently.
Table~\ref{M-tab:budget} and Table~\ref{tab:smobs} collect the outcome at the three radii.

\emph{Solver step.} On the base mesh the solver step is halved three times, from $2\pi/96$ to
$2\pi/768$. The Eulerian and tracer means converge at observed orders between 0.93 and 0.96,
consistent with the old-time-level explicit terms, and this step error accounts for most of the base
departures in Table~\ref{M-tab:budget}.

\emph{Tracer integration.} The base tracers take Heun steps equal to the solver step, with velocities
interpolated linearly in time between the two time-bound snapshots. Four Heun substeps with velocities
interpolated quadratically through three snapshots change $U_L$ by at most 0.046 per cent at the base
step and 0.012 per cent at $2\pi/192$. Refining only the Heun substeps, with linear interpolation,
does not remove this interpolation error: four linear substeps miss the converged value by the same
amount as the base integration, with the opposite sign.

\emph{Grid, startup and window.} At solver step $2\pi/384$, $144$ instead of $96$ radial nodes change
$U_E$ and $U_L$ by at most 0.035 and 0.046 per cent, and $36$ instead of $24$ angular points by at most
$1.4\times10^{-8}$ and $2.1\times10^{-8}$ per cent. At the same step, $64$ radial nodes as a third grid give
observed spatial orders of 1.93 to 2.02 for $U_E$ and for $U_L$ of both releases,
consistent with the second-order rows near the grid ends. Releasing the tracers after six instead of three periods changes $U_L$ by at
most $1.2\times10^{-11}$ per cent, and a six-period window by at most $4.7\times10^{-4}$ per cent.
Full evolved-state comparisons at every solver phase measure numerical periodicity: on the base mesh
after nine periods the relative area-weighted discrete RMS recurrence residual is $4.62\times10^{-7}$ for
vorticity and at most $1.14\times10^{-9}$ for velocity.

\emph{The tracer observable.} The coefficients refer to reference positions, whereas the base observable
releases tracers on a physical ring of radius $r$ and reports $r$ times their mean angular velocity.
At $r=0.60$ these tracers' mean radii exceed $r$ by $1.71\varepsilon^2$ on average
and have a standard deviation of $0.69\varepsilon$ around the ring.
The mapped release starts each tracer at the first-order displaced position of a reference
point $\bm X$ on the ring, applying the analytic $\bm\xi_1(\bm X,\tau_0)$ in polar components,
$r\mapsto r+\xi_{1r}$ and $\theta\mapsto\theta+\xi_{1\theta}/r$;
its mean radius equals $r$ to within $2.7\times10^{-3}\,\varepsilon^2$.
The mean-radius comparison instead uses the same physical-ring tracers and evaluates
the leading coefficient at each tracer's own mean radius. For the unit travelling wave,
write $\mathcal L(r)=E_m(r)+G(r)$. This comparison averages the individual ratios,
\begin{equation}
\delta_{L,\bar r}=\frac{1}{N_p}\sum_{i=1}^{N_p}
\frac{\bar r_i\,\Delta\theta_i}{T\varepsilon^2\mathcal L(\bar r_i)}-1,
\label{eq:meanradius}
\end{equation}
where $N_p=16$, $\bar r_i$ is the time-averaged radius, and $T$ is the duration
in units of $1/\omega$. All three observables use the difference $\Delta\theta_i$
of each tracer's unwrapped angle at the two endpoints of the measurement window,
which have the same wall phase. Thus correcting the reference radius retains the
endpoint estimator.

\begin{table}[t]
\caption{Tracer-observable comparisons at $m=2$, $\mathrm{Wo}^2=10$, complementing Table~\ref{M-tab:budget}. Here
$D=[1-U_L/(U_E+U_S)]/\varepsilon^2$, and $c_L$ multiplies $\varepsilon^2$ in $\delta_L$. Mean radius: the per-tracer
ratios in Eq.~\eqref{eq:meanradius}. Fitting $1-U_L/(U_E+U_S)=n+D_0\varepsilon^2$ separates the amplitude-independent
numerical part $n$ from $D_0$, the limit of $D$ after that numerical part is removed. The fits are those of
Table~\ref{M-tab:budget}. The base row uses the base tracer integration of Sec.~\ref{sec:smprotocol}; the other rows use
four Heun substeps with quadratic interpolation in time.}
\label{tab:smobs}
{\small\noindent\textit{Alt text.} The physical-ring release gives a larger amplitude correction
than the mapped release at all three radii. The difference grows towards the wall.\par}
\begin{ruledtabular}
\begin{tabular}{lccc}
 & $r=0.45$ & $r=0.60$ & $r=0.75$\\
\hline
Base, $\varepsilon=0.02$: $D$ & $+1.95$ & $+12.28$ & $+19.87$\\
$c_L$, mean radius & $+5.05$ & $+0.28$ & $-1.57$\\
$D_0$, physical ring & $+3.3$ & $+12.3$ & $+16.1$\\
$D_0$, mapped release & $-0.01$ & $+2.90$ & $+4.23$\\
\end{tabular}
\end{ruledtabular}
\end{table}

\section{Amplitude separation and tracer observables}
\label{sec:smamp}
To separate numerical and finite-amplitude effects, we fit the base departures at five
amplitudes between 0.01 and 0.04, including 0.015, 0.02 and 0.03, at fixed resolution.
The fit assigns an amplitude-independent offset to numerical error and a term proportional to
$\varepsilon^2$ to finite amplitude. It reproduces $\delta_L$ and $D$ to within $0.45$ and $0.57$
in units of $\varepsilon^2$, with the largest residuals at the smallest amplitude and
at most $0.10$ at the three intermediate amplitudes. In units of the last digit printed in
Tables~\ref{M-tab:budget} and~\ref{tab:smobs}, the largest residual is 2 for $c_E$, 4 and 6 for the physical-ring $c_L$ and $D_0$,
up to 7 for the mapped-release rows and 13 for the mean-radius row.

Extrapolated to zero solver step, the numerical part is at most 0.07 per cent for $\delta_E$,
0.09 per cent for $\delta_L$ of the mapped release and 0.10 per cent for the physical ring
(Table~\ref{M-tab:budget}), with extrapolation uncertainties of at most 0.10 and 0.15 per cent.
For each quantity, the uncertainty is the larger of the absolute difference between the finest-step value and the
extrapolate and the change under extrapolation from the preceding three step sizes. Removing
the spatial error using 64, 96 and 144 radial nodes (observed orders 1.93 to 2.02)
leaves at most 0.007 and 0.015 per cent for the Eulerian and tracer means, respectively,
for either tracer release. The time-extrapolation uncertainty limits the comparison. The moving-domain computation
reproduces the leading Eulerian and total coefficients at the three radii to within
0.11 and 0.17 per cent when the absolute remaining numerical part and this uncertainty
are added. These empirical bounds concern the amplitude-independent part. We round upper bounds upwards.

At $\varepsilon=0.02$, time extrapolation at finite amplitude gives departures of
$+0.04$, $+0.09$ and $+0.10$ per cent for the Eulerian mean at the three radii in table order.
The tracer departures are $+0.03$, $-0.07$ and $-0.09$ per cent for the mapped release
and $-0.10$, $-0.45$ and $-0.58$ per cent for the physical-ring release.

The $\varepsilon^2$ parts are finite-amplitude effects. For the Eulerian mean at fixed physical points the
coefficient is $c_E=2.45$ to $2.64$ and changes by less than 0.01 on $144$ radial nodes. For the
tracer drift it depends on the observable: at $r=0.60$, $c_L$ is $-10.3$ for physical-ring release,
$-0.91$ for the mapped release and $+0.28$ at each tracer's mean radius, and these differ by up to 12.0
at the other radii. The mean-radius comparison uses the individual ratios in
Eq.~\eqref{eq:meanradius}, so subtracting its fitted coefficient from the physical-ring
coefficient does not isolate an endpoint correction.
The endpoint angle difference samples each tracer's first-order oscillation at a wave
phase advanced by its own drift. For physical-ring release this sampling correlates with
the tracers' mean-radius offsets and mean phases. A first-order evaluation gives
$+1.1$, $-3.3$ and $-7.9$ in the physical-ring $c_L$ at the three radii.
This contribution vanishes at that order for the mapped release; averaging each tracer's
angle over one period at each end of the window removes the first-order sampling contribution.

At $r=0.60$, the fits at $\mathrm{Wo}^2=10$ give $D_0=12.3$ for the physical ring and $2.90$ for the
mapped release. Neither limit changes by more than 0.01 on 144 radial nodes.

The dependence on release position also appears in creeping flow, where the instantaneous
velocity field solves the Stokes problem in the deformed disk. For a single travelling wave the domain and its wall velocity rotate at angular
speed $1/m$, so the field at every phase is the rotated field at $\tau=0$.
A truncated expansion in biharmonic modes, up to angular order 64, satisfies the wall data by
collocation with a residual below $10^{-13}$, and adaptive tracer integration at relative tolerance
$10^{-13}$ in the phase-rotated velocity field gives at $r=0.60$ the limits
$D_0=12.55$ for physical-ring release and $2.51$ for mapped release; $D$ changes by less than
$10^{-9}$ between 40, 64 and 96 modes.
At $\varepsilon=0.02$ the same creeping-flow calculation gives $D=12.53$ for the physical ring;
the coarse moving-domain sweep at that amplitude gives $D=12.72$ to $12.74$ at
$\mathrm{Wo}^2=0.5$ to $2$. The difference is about $8\times10^{-5}$ in
$U_L/(U_E+U_S)$ and includes both the departure from the viscous limit and the
coarse discretization.

The fitted $c_E$ is the same at every radius and, extrapolated from
$\varepsilon=0.0025$ and $0.005$, agrees with $(6+3m-2m^2)/2$ to six digits for $m=1$ to $6$,
giving $2$ at $m=2$. An independent implementation of the same creeping-flow calculation gives
the tracer drift for $m=1$, $2$ and $3$ at the three radii: extrapolated to $\varepsilon\to0$ it
agrees with Eq.~\eqref{M-eq:Lsmall} within $10^{-8}$, and at $m=1$ it opposes the wave at every
amplitude from $\varepsilon=0.005$ to $0.04$. For $m=2$ with tangential wall motion
($\gamma=\pm1/2$), the physical-ring drifts at $r=0.45$, $0.60$ and $0.75$ were
extrapolated from the same four amplitudes. Their leading coefficients agree with
Eq.~\eqref{M-eq:gamma12} to within $8\times10^{-9}$ in absolute value.

\section{Controls at small Womersley number and for standing and mixed waves}
\label{sec:smcontrols}
The standing and mixed controls prescribe
\begin{equation}
R^2=1-2\varepsilon[A\sin(m\phi-\tau)+B\sin(m\phi+\tau)],
\label{eq:controlwall}
\end{equation}
with $a=-iA$ and $b=iB$. The standing wall has $A=B=1/2$, giving
$R^2=1-2\varepsilon\sin(m\phi)\cos\tau$.
This phase convention has the same ring averages as the real-amplitude standing mode.
The mixed wall has $A=4/5$, $B=3/5$, $S_0=A^2+B^2=1$ and $S_3=A^2-B^2=7/25$.
Its peak first-order wall velocity is $1.4\varepsilon\omega R_0$, so matching $S_0$
matches intensity while increasing the peak velocity relative to the travelling control.

On the coarse sweep, at $\mathrm{Wo}^2=0.5$ and $1$, the measured Eulerian mean at
$r=0.60$ is within $0.06$ per cent of Eq.~\eqref{M-eq:Esmall} and $0.05$ per cent of
Eq.~\eqref{M-eq:Efinite}. The total drift reconstructed from the recorded decomposition ratio is
within $0.55$ per cent of both Eq.~\eqref{M-eq:Lsmall} and Eq.~\eqref{M-eq:Lfinite}; these coarse runs use
physical-ring release. For scale, the finite-amplitude contribution to this tracer observable is
$-0.41$ per cent at $\mathrm{Wo}^2=10$ and $\varepsilon=0.02$ (Table~\ref{M-tab:budget}).
At other values of $\mathrm{Wo}$ this contribution depends on the first-order profile.

At the same radius the analytic ratio $U_E/U_S$ is 4.3675 at
$\mathrm{Wo}^2=0.5$ and 3.3838 at $\mathrm{Wo}^2=10$,
compared with coarse measured ratios $4.3657$ and $3.4199$.

The Stokes drift on this ring is particularly sensitive to $\mathrm{Wo}$ for $m=2$.
As $\mathrm{Wo}^2$ rises from 2 to 10, its departure from the viscous-limit coefficient grows
from 0.87 to 20.55 per cent. The corresponding Eulerian departure grows from 0.28 to 6.67 per cent.
The total drift stays closer to its viscous-limit value, with departures of 0.07 and 1.60 per cent.
Each percentage is relative to the viscous-limit coefficient for that quantity.
The base comparison uses the coefficients at $\mathrm{Wo}^2=10$.

The standing Eulerian ring mean falls to $7\times10^{-15}$ of its travelling value;
the standing tracer value at $\varepsilon=0.02$ is $-1.06\times10^{-17}\omega R_0$, or
$-1.82\times10^{-14}$ times the travelling value at $r=0.60$.
These cancellations test reflection symmetry of ring averages.
On the coarse mixed-wall control, $U_L$ differs from $7/25$ of the travelling value
by $0.240$ per cent. This comparison tests the leading chirality rule at finite
amplitude and at the coarse solver and tracer resolution.
The departure is smaller than the $0.49$ per cent by which the coarse travelling control itself
differs from $U_E+U_S$, so the comparison supports the rule at that level and does not resolve it more finely.

\section{Finite-element implementation}
\label{sec:smfe}
A curved P3/P2 finite-element code computes the first-order field on the reference disk,
transfers the second-order boundary condition, and constructs the Stokes drift and the total drift.
The eight cases are $m=2,3$ at $\mathrm{Wo}^2=10^{-4},1,10,100$. The zeros of its ring-averaged
coefficients converge to the Bessel evaluation under mesh refinement: over three refinement levels the
largest Stokes-drift zero error falls from $2.4\times10^{-6}$ to $6.1\times10^{-8}$, and the near-wall zero of the total drift at $m=2$,
$\mathrm{Wo}^2=100$ is approached to within $2.8\times10^{-4}$, $2.3\times10^{-5}$ and $2.5\times10^{-6}$. The total drift stays positive at
the Stokes-drift zeros. On fixed-radius rings the angular content of the computed drift is below
$4.8\times10^{-6}$ of the ring mean on the finest mesh and lies entirely at harmonics of the mesh's fourfold
symmetry.

The near-wall zero provides a sensitive test. Radial refinement from 129 to 4097 nodes and
high-precision quadrature agree on its position at $m=2$ and $\mathrm{Wo}^2=100$.
At $r=0.98$ the Eulerian and Stokes contributions, $3.803170$ and $-3.827768$,
leave a total coefficient of $-0.024598$. On the sampled profile the minimum is $-0.027$
at $r=0.985$ and the positive peak is 1.01 at $r=0.73$.

\section{Wall with tangential motion}
\label{sec:smgamma}
A material point labelled $\phi$ moves to $\theta_w=\phi-\varepsilon\gamma\cos(m\phi-\tau)$
on the prescribed shape $R^2=1-2\varepsilon\sin(m\theta-\tau)$. Its first-order radial and
tangential displacements are $-\varepsilon\sin(m\phi-\tau)$ and
$-\varepsilon\gamma\cos(m\phi-\tau)$ (Sec.~\ref{M-sec:total}). The solver adds
$\partial_\tau\theta_w\,\partial_\theta(R\bm e_r)$ to the radial-wall velocity,
with $\partial_\theta(R\bm e_r)=R_\theta\bm e_r+R\bm e_\theta$.
This tangential motion preserves the normal flux and the radial-wall streamfunction.
At each grid angle the azimuthal wall velocity is $R\,\partial_\tau\theta_w$ for the
corresponding material label. The computed wall velocity agrees with the prescribed velocity
to within $10^{-11}\omega R_0$ at the end of every period. At $\gamma=0$ the Eulerian ring means
and physical-ring tracer angles reproduce the radial-wall results of Sec.~\ref{sec:smprotocol} bit for bit.

For $m=2$, $\mathrm{Wo}^2=10$ and $\gamma=1/2$, we use the radial-wall base mesh
and double the time resolution from 96 to 768 steps per period at $\varepsilon=0.02$.
At 192 and 384 steps we also test $\varepsilon=0.01$ and $0.04$. Mapped release
uses the first-order displacement for $f'(1)=1/2$. The observed time orders lie between 0.93 and 0.96.
At 384 steps and $\varepsilon=0.02$, refinement from 96 to 144 radial nodes changes the result by
at most 0.03 per cent in $U_E$ and 0.02 per cent in $U_L$. The three-amplitude fits
leave residuals of at most $0.12\varepsilon^2$.

The extrapolated numerical part is at most 0.07 per cent for $\delta_E$ and 0.04 per cent for $\delta_L$,
with Richardson uncertainties of at most 0.14 and 0.10 per cent.
At each radius we add the absolute numerical part to its uncertainty (Sec.~\ref{sec:smamp})
and round the bound upwards (Table~\ref{tab:smgamma}). The Eulerian bound reaches 0.20 per cent
at $r=0.45$ and the tracer bound reaches 0.13 per cent at $r=0.75$ for both releases.

\begin{table}[b]
\caption{Moving-domain departures for the first-order inextensible wall at $m=2$ and $\mathrm{Wo}^2=10$.
Notation as in Table~\ref{M-tab:budget}. Equations~\eqref{M-eq:Efinite} and \eqref{M-eq:Lfinite}
give $U^{(0)}$ with $f'(1)=1/2$. The coefficients $c$ use all three amplitudes at 384 steps per period.}
\label{tab:smgamma}
{\small\noindent\textit{Alt text.} Across the three rings, the largest combined error bound is
0.20 per cent for the Eulerian mean and 0.13 per cent for either tracer release.\par}
\begin{ruledtabular}
\begin{tabular}{lccc}
 & $r=0.45$ & $r=0.60$ & $r=0.75$\\
\hline
Base, $\varepsilon=0.02$: $100\delta_E$ & $+1.24$ & $+1.34$ & $+1.29$\\
\quad $100\delta_L$, physical ring & $+0.48$ & $+0.35$ & $+0.32$\\
Numerical, $\Delta\tau\to0$: $100\delta_E$ & $-0.06$ & $-0.01$ & $+0.01$\\
\quad $100\delta_L$, physical ring & $-0.01$ & $-0.01$ & $-0.04$\\
\quad $100\delta_L$, mapped release & $-0.01$ & $-0.01$ & $-0.04$\\
Combined bound, $100\delta_E$ & $0.20$ & $0.15$ & $0.13$\\
\quad $100\delta_L$, physical ring & $0.05$ & $0.07$ & $0.13$\\
\quad $100\delta_L$, mapped release & $0.06$ & $0.08$ & $0.13$\\
$c_E$, fixed points & $+9.8$ & $+10.1$ & $+10.3$\\
$c_L$, physical ring & $+3.5$ & $-2.7$ & $-8.1$\\
$c_L$, mapped release & $+4.0$ & $+1.3$ & $+0.4$\\
\end{tabular}
\end{ruledtabular}
\end{table}